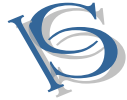

IJCSBI.ORG

# Digital Image Tamper Detection Techniques - A Comprehensive Study

**Minati Mishra**
Department of Information and Communication Technology
Fakir Mohan University, Balasore, Odisha, India

**Flt. Lt. Dr. M. C. Adhikary**
Department of Applied Physics and Ballistics
Fakir Mohan University, Balasore, Odisha, India

## ABSTRACT

Photographs are considered to be the most powerful and trustworthy media of expression. For a long time, those were accepted as proves of evidences in varied fields such as journalism, forensic investigations, military intelligence, scientific research and publications, crime detection and legal proceedings, investigation of insurance claims, medical imaging etc. Today, digital images have completely replaced the conventional photographs from every sphere of life but unfortunately, they seldom enjoy the credibility of their conventional counterparts, thanks to the rapid advancements in the field of digital image processing. The increasing availability of low cost and sometimes free of cost image editing software such as Photoshop, Corel Paint Shop, Photoscape, PhotoPlus, GIMP and Pixelmator have made the tampering of digital images even more easier and a common practice. Now it has become quite impossible to say whether a photograph is a genuine camera output or a manipulated version of it just by looking at it. As a result, photographs have almost lost their reliability and place as proves of evidences in all fields. This is why digital image tamper detection has emerged as an important research area to establish the authenticity of digital photographs by separating the tampered lots from the original ones. This paper gives a brief history of image tampering and a state-of-the-art review of the tamper detection techniques.

## 1. INTRODUCTION

Today, digital images not only provide forged information but also work as agents of secret communication. Users and editing specialists manipulate digital images with varied goals. Scientists and researchers manipulate images for their work to get published; medical images are tampered to misrepresent the patients' diagnostics, photo and yellow journalists use the trick for creating and giving dramatic effect to their stories, politicians, lawyers, forensic investigators use tampered images to direct the opinion of people, court or law to their favour and so on. Hence, distinguishing the original images from faked lots and establishing the authenticity of digital photographs have become some of the greatest challenges of the present time. Retouching, splicing, copy-pasting, cropping, cloning etc are some of the popular techniques used for image manipulations. In additions to these techniques there also exists a wide range of Steganographic methods those use this popular digital media for secret data transmission.

This paper is organized as follows. A brief history of Image tampering is given in the next section. Classification of tampering techniques and distinction between image tampering and Steganography is made in Section 3. A state-of-the-art review of the existing tamper detection techniques is given in section 4 which is followed by the summary and conclusion at the end.

## 2. HISTORY OF IMAGE TAMPERING

According to Oxford dictionary, the literary meaning of 'tampering' is interfering with something so as to make unauthorized alterations or damages to it [1]. Though it emerged as a critical problem with the availability of digital cameras but Photo fakery is not an issue of 21st century. Its history can be dated back to as early as in 1840s. Hippolyte Bayard, the first person to create a fake image as recorded by history, is famous for a picture of him committing suicide (Figure. 1). It was later found that the photograph was forged out of frustration because he had lost the chance of becoming 'the inventor of photography' to Louis Daguerre. Daguerre patented a photography process earlier than him and owned all the glory [2].

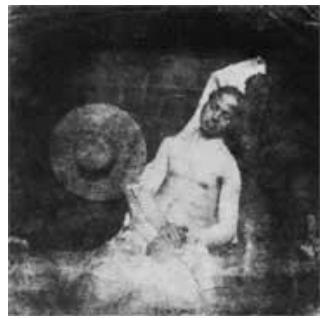

**Figure 1. Hippolyte Bayard's suicidal photograph**

Before computers, photo manipulations were performed by retouching with ink, paint, by double-exposing, airbrushing piecing photos or negatives together in the darkroom, or scratching Polaroid. In the early days of photography, the use of technology was not as advanced and efficient as it is now. The outputs of such manipulations were very much similar to digital manipulations but they were more difficult to create [3].

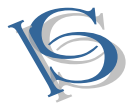

## 2.1 Conventional Vs Digital Image Tampering

Digital Image tampering is very much similar in nature to that of the conventional photo fakeries where the conventional photographs are replaced by their digital counterparts. One of the key characteristics of a digital image is; it is easier to modify or manipulate a digital image in comparison to its conventional counterpart. The process of digital image tampering have been made even more simple and easier, thanks to the availability of low cost (or sometimes free of cost) and powerful image editing software nowadays.

History of digital image tampering can be dated back to the late twentieth century to support political propagandas [3]. According to the Wall Street Journal, even in 1989, out of all the colour photographs published in United States, 10% were digitally altered or retouched! [4] An article in the journal Nature reports on the impact of digital photography and image-manipulation software in science. Mike Rossner, editor of the Journal of Cell Biology, estimates that roughly 20% of accepted manuscripts to his journal contain at least one figure that has to be remade because of inappropriate image manipulation. And, in 1990, 2.5% of allegations examined by the U.S. Office of Research Integrity, which monitors scientific misconduct, involved contested scientific images. By 2001, this figure was nearly 26% [5]. We suspect, as on date, more than 90% of commercially captured and published digital photographs, if not tampered with some ulterior motive, are at least retouched to improve the look, colour, contrast or background.

Few of the well-known historical tampered images include: The iconic portraits of U.S President Abraham Lincoln (Figure. 2c) and Benito Mussolini (Figure. 2d), the Iraq soldier picture of the Los Angeles Times (March 1, 2004), the Internet image of a tourist just before the 911 WTC attack (Sept 2001) (Figure. 2b) and that showing Jane Fonda and John Kerry sharing the same speaker platform (Feb 2004) (Figure. 2a). Coming to Indian context, the sensational photograph that sent the Congress MLA, Kalpana Parulekar into police custody on $3^{rd}$ February, 2012 was a tampered photograph of State Lokayukta P P Naolekar. She has used that photo to show that Naolekar was a former RSS worker (Figure. 2e). The former two photographs given here represent the conventional photo fakeries whereas the later are the results of sophisticated digital image tampering.

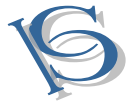

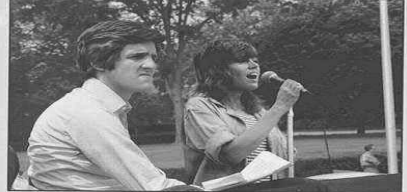

**Figure 2a.** This image was released during the 2004 presidential election campaign which shows John Kerry and Jane Fonda speaking together at an anti-Vietnam war protest which later turned out to be a politically motivated forged composite of two different images of Kerry taken on June 13, 1971 at an anti-war rally in Mineola, N.Y. and another of Jane Fonda taken in August, 1972 speaking at a Political rally at Miami [6] [7].

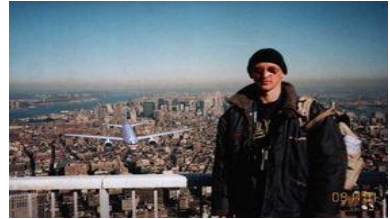

**Figure 2b.** This famous altered image smacked emails and went viral after the 911 attacks in New York. The photo was forwarded with the title "the tourist guy, the accidental tourist, Waldo or the WTC Guy" and displayed a person standing on the Observation deck of one of the World Trade Centre towers seconds before the plane hit the tower. Later it was found to be an edited image of Péter Guzli, a 25 year old Hungarian man taken during his visit to relatives in N. Y. on 28/11/1997 [8].

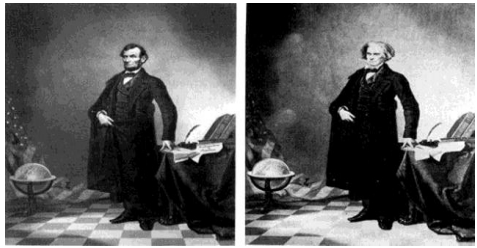

**Figure 2c.** This iconic portrait of U.S President Lincoln is a composite of Lincoln's head and the Southern politician John Calhoun's body [9].

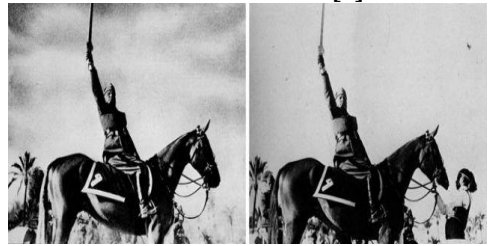

**Figure 2d.** The horse handler had been airbrushed from the original photograph of Benito Mussolini to make him look more epic and heroic [9].

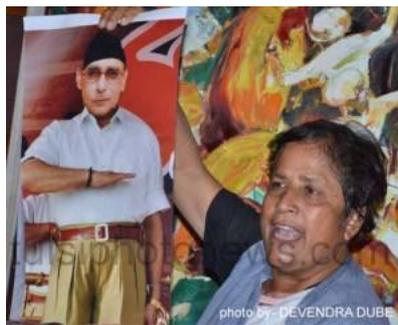

**Figure 2e.** Congress MLA Kalpana Parulekar showing a doctored photograph of State Lokayukta P P Naolekar to show that he was a former RSS worker.[10]

## 3. YPES OF DIGITAL IMAGE TAMPERING

There are many more such cases of digital image tampering available and the list is increasing every second with addition of newer cases. Based on whether the manipulation is performed to the visible surface of the image or

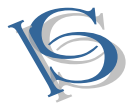

IJCSBI.ORG

to invisible planes, the manipulation techniques can be classified as image tampering or image Steganography. Image tampering again can be performed either by making changes to the context of the scene elements or without the change of the context. In the second case, the recipient is duped to believe that the objects in an image are something else from what they really are but the image itself is not altered [11]. Figure.3 shows such an image published in November 1997 after 58 tourists were killed in a terrorist attack at the temple of Hatshepsut in Luxor Egypt in which a puddle of water were digitally altered to appear as blood flowing from the temple [12].

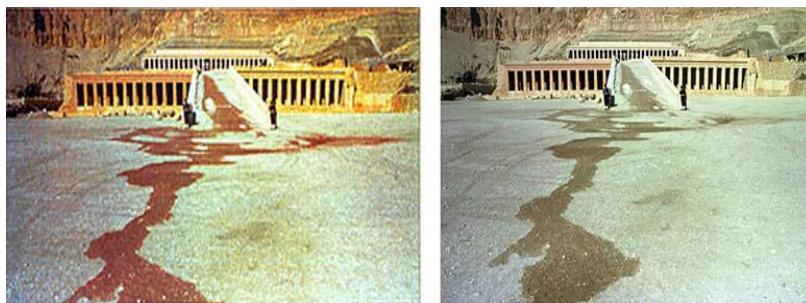

**Figure 3. A digitally altered photograph (left) of temple of Hatshepsut (right) after a terrorist attack.**

The context based image tampering methods, according to the manipulation process used, are further divided into three major classes such as: Retouching, Splicing, and Copy-Move or Cloning. These manipulations generally involve deletion or addition of scene elements of an image or combination of scenes from multiple images [13]. Besides these above mentioned techniques there is another class of photo fakery known as Computer Graphics created photography. Though it requires a little more effort still by the use sophisticated computer graphics rendering software highly photorealistic images are created which are impossible to be differentiated from the photographic images [14] [15] [16]. Figure.4 shows two such photorealistic images.

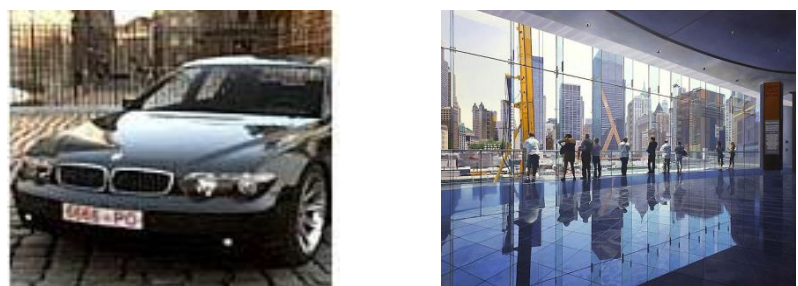

**Figure 4. Computer Graphics Created photorealistic image (left) source: [12] and a photorealistic painting (right) source: [17]**

## 3.1 Digital Image Tampering Vs Digital Image Steganography

Tampering and Steganography, though both the techniques manipulate a digital image from its original capture but, they differ from each other at their vary purposes. One manipulates an image for the purpose of hidden communication whereas the other manipulates it to fake a fact and mislead the viewer to misbelieve the truth behind a scene [18][19].

Image Steganography is the process of secret communication where a piece of information (a secret message or an image, preferably encrypted) is encoded into the bits of an innocent looking cover image, in such a manner that the very existence of the secret information remains concealed without raising any suspicion in the minds of the viewers [20] whereas, image tampering generally replaces one or more parts of a host image with those from the same or from some different images [21].

Because of its inherent purpose of data hiding, Steganography requires the original and the Stego image to look alike but tampering on the other hand aims at creating an image that appears to be an original camera output by copy pasting parts from one or more images and preferably applying geometric transformations such as scaling, rotation, cropping as well as smoothing operations such as edge blurring, blending etc. to the copied parts to ensure the tampered image to look as natural as possible and the tampering undetectable to human vision.

Steganography is more global in nature and offer vary little or no change to the image content in comparison to tampering which makes dramatic changes to the image content those are more local in nature[21].

Figures.2c and 2d show two different tampered images and their original counterparts and figure.5, given below, shows a Stego Lena image that has been created by embedding a piece of hidden information into the original Lena image. The differences between the pairs of tampered images and their original ones are clearly visible where as there exists no visible difference between both the Lena images.

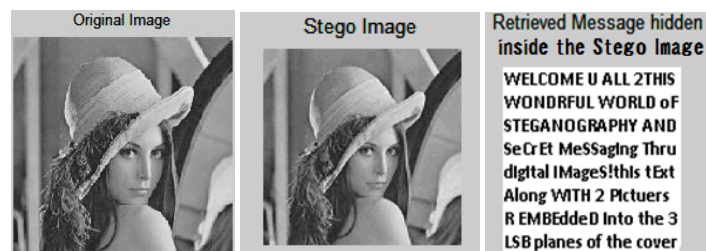

**Figure 5. Original Lena image, Stego Lena and the Retrieved hidden Information**

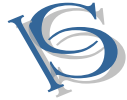

## 4. TAMPER DETECTION TECHNIQUES

Digital image tamper detection techniques can be broadly classified into two groups such as active detection techniques and passive (blind) techniques. The active techniques require a pre-processing step and suggest embedding of watermarks or digital signatures to images so as to authenticate them. The major difficulty with this method is that it requires the watermark to be embedded at the time of image capturing and for this; all digital cameras should have a standard inbuilt watermark. Few questions need to be answered in this regard are: whether all the camera manufacturing companies will agree to manufacture cameras with some standard watermarks signals inbuilt into them? Whether the costumers will be ready to accept the probable degradation in the image quality due to the embedded watermark? What about the processing time and complexity that involves the embedding and retrieval of the watermark? Most importantly how to deal with all those millions of pre manufactured digital cameras already available in market as well as with users and can false watermarking be completely ruled out? All these questions make the image authentication and active tamper detection technique a remote possibility in practice. On the other hand, the passive detection techniques do not require pre embedding of any watermark or digital signatures to the images and hence are commonly used for the purpose of tamper detection in digital images.

### 4.1 Active Methods of Tamper Detection

Active taper detection techniques due to their inherent limitation, though, are not as common as those of the passive techniques still these are considered to be most efficient image authentication methods and a lot of research has been done in this field. These active image authentication techniques are commonly classified into two categories: the first method uses a fragile watermark, which localizes and detects the modifications to the contents. While the rate of tamper detection is very high for these methods they cannot distinguish between the simple brightness, contrast adjustments and replacement or addition of scene elements. Increasing the gray scales of all pixels by one would indicate a large extent of tampering by this method, even though the image content remains unchanged for all practical purposes [23]. The second method uses a semi-fragile watermarking, that only detects the significant changes in the image while permitting content-preserving processing.

The fragile watermark though has good localization and security properties but cannot differentiate forgeries such as addition or removal of parts of image, from the innocent image processing operations such as brightness or contrast adjustments. J. Fridrich [23] solves this problem through his new hybrid image authentication watermarking scheme that combines both the fragile and a robust watermark. The hybrid watermark can be used to accurately pinpoint changes as well as distinguish forgeries from other innocent operations. This work is further improved and a secured hybrid method [24] is presented in by Deguillaume and Voloshynovskiy. Several researchers worked in these active tamper detection and authentication schemes and developed a number of fragile, semi-fragile, robust, public as well as private key based watermarks for copyright protection, authentication and tamper detection [25-29] out of which, some either failed to effectively address the problems or sacrifice tamper localization accuracy of the original methods while few of them were proved to be highly efficient and effective. However, the hierarchical digital watermarking method proposed by Phen et.al is a simple but efficient method that not only localizes and detects tampering but also is capable of tamper recovery with a little degradation to the image quality. The precision of tamper detection and localization of this method is 99.6% and 100% after level-2 and level-3 inspection, respectively. The tamper recovery rate is better than 93% for a less than half tampered image [22].

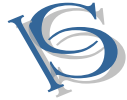

## 4.2 Passive Detection Techniques

The passive methods are regarded as evolutionary developments in the area of tamper detection. In contrast to the active authentication techniques these methods neither require any prior information about the image nor necessitate the pre embedding of any watermark or digital signature into the image. The underlying assumption that is the basis of these schemes is, though the carefully performed digital forgeries do not leave any visual clue of alteration, they are bound to alter the statistical properties of the image. The passive techniques try to detect digital tampering in the absence the original photograph as well as without any pre inserted watermark just by studying the statistical variations of the images [30]. Researchers of passive detection techniques generally focus on two types of passive methods, the copy-move forgery detection or cloning and splicing.

### 4.2.1 Cloning Detection

To clone or copy and paste a part of the image to conceal an object or person is one of the most commonly used image manipulation techniques. When it is done with care, it becomes almost impossible to detect the clone visually and since the cloned region can be of any shape and size and can be located anywhere in the image, it is not computationally possible to make an exhaustive search of all sizes to all possible image locations.

According to [31], any Copy-Move forgery introduces a correlation between the original image segment and the pasted one which can be used as a basis for successful detection of this type of forgeries. Because the tampered image will likely be compressed and because of a probable use of the smoothing or other post processing operation, the segments may only match approximately not exactly. The authors in this paper give two different detection schemes: exact and robust matching those successfully detects duplicate regions in an image even when the images are post processed following a copy-paste. Methods based on blur movement invariants and DWT, SVD, PCA based sorted neighbourhood approaches are suggested in [32][33][34] for robust detection of cloned regions in an image.

*4.2.2 Splicing Detection Techniques*
Digital splicing of two or more images into a single image is another commonly used image manipulation technique. When performed carefully, the borders between the spliced regions can be visually imperceptible. It is a popular way to distort the semantic content of an image so as to fool the viewer to misbelieve the truth behind a scene. Image splicing is a fundamental operation in image forgery and is characterized by simple cut-and-paste operation that takes a part of an image and puts it onto either the same or another image without performing any post-processing smoothing operation such as edge blurring, blending to it. By Image tampering, it generally means splicing followed by the post-processing operations so as to make the manipulation imperceptible to human vision [18].

Splicing detection is more challenging in comparison to cloning detection as unlike cloned images spliced images do not have any duplicate regions and unavailability of the source images offer no clue about the forgery. In [31], however, the authors have shown that splicing disrupts higher-order Fourier statistics, which can subsequently be used to detect splicing. Tian-Tsong Ng and Shih-Fu Chang in [35] suggest a bio- coherence feature based splicing model. Yun Q. Shi, Chunhua Chen, Wen Chen in [36] proposed an effective splicing detection approach based on a natural image model that consists of statistical features extracted from the given test image as well as 2-D arrays generated by applying multi-size block DCT transform to the test images. With the assumption that fusion of multiple statistical features can improve the performance of splicing detection, Jing Zhang, Yun Zhao, Yuting Su in their paper proposed a new splicing detection approach based on the features utilized for steganalysis. They merge Markov process based features and DCT features for splicing detection. The proposed approach achieved up to 91.5% accuracy with a 109-dimensional feature vector [37]. In [38] the authors proposed an automatic detection framework to identify a spliced image based on a human visual system (HVS) model in which visual saliency and fixation are used to guide the feature extraction mechanism. Zimba and Xingming in their paper propose a novel method for detecting image splicing by thresholding transition region measures of DWT coefficients of a suspicious image in chroma spaces. Only the low frequency sub-band of the DWT of the suspected image is extracted to reduce the size of the image and improve the performance [39]. Because splicing combines image parts from multiple images so, careful study of the lighting conditions can provide a better clue on detection of these types of manipulations.

## 5. SUMMARY AND CONCLUSIONS

A lot of research has been done on active as well as passive tamper detection techniques and still a lot of work is going on worldwide to successfully detect tampering in digital images. In this paper we have reviewed the two popularly used passive detection techniques, splicing and cloning. There exits many other techniques such as detection based on examining the lighting environment, camera feature based detections, studying the statistical and geometric properties. But interestingly the tamper detection techniques and the tampering techniques are evolving together. With the development of a new image authentication and tamper detection technique, newer variants of detection resistant tampering methods are evolving making the issue more challenging.